\documentclass[aps,prl,showpacs,twocolumn]{revtex4}%
\usepackage{amsfonts}
\usepackage{amssymb}
\usepackage{amsmath}
\usepackage{graphicx}
\usepackage{dcolumn}
\usepackage{bm}
\usepackage{units}
\usepackage{color}%

\begin{document}

\title{Shot noise in magnetic tunnel junctions from first principles}
\author{Kai Liu}
\affiliation{Department of Physics, Beijing Normal University, Beijing 100875, China}
\author{Ke Xia}
\affiliation{Department of Physics, Beijing Normal University, Beijing 100875, China}
\author{Gerrit E. W. Bauer}
\affiliation{Institute for Materials Research, Tohoku University, Sendai 980-8577, Japan}
\affiliation{Kavli Institute of NanoScience, Delft University of Technology, 2628 CJ
Delft, The Netherlands}
\date{\today }

\begin{abstract}
We compute the shot noise in ballistic and disordered Fe$|$MgO$|$Fe tunnel
junctions by a wave function-matching method. For tunnel barriers with $%
\lesssim $5 atomic layers we find a suppression of the Fano factor as a
function of the magnetic configuration. For thicker MgO barriers the shot
noise is suppressed up to a threshold bias indicating the onset of resonant
tunneling. We find excellent agreement with recent experiments when
interface disorder is taken into account
\end{abstract}

\pacs{72.25.Ba, 85.75.-d, 72.10.Bg }
\maketitle

The statistics of electron transport in mesoscopic systems has been subject
to intensive research in the last decades leading to important and useful
insights \cite{Blanter, QN}. In a two-terminal conductor with a
time-dependent current $I(t)$, the simplest measure is the noise power $%
P(\omega )=\int_{-\infty }^{\infty }\langle \triangle I(0)\triangle
I(t)\rangle e^{i\omega t}dt$, where $\triangle I(t)\equiv I(t)-\langle
I\rangle $ denote the instantaneous fluctuation from the average current and
$\left\langle \cdots \right\rangle $ a time and statistical average. The
shot noise $S$ is the zero frequency limit of the noise power when the
applied voltage $\left\vert eV\right\vert $ is sufficiently larger than the
thermal energy $k_{B}T.$ The classical shot noise characterized by an
uncorrelated Poissonian process is given by the Schottky formula $%
S=2e\langle I\rangle $ \cite{Schottky}. Shot noise contains information
about the charge of the elementary excitations, entanglement, wave \textit{vs%
}. particle nature of electron transport, and provides a diagnostic for open
transport channels \cite{BeenSchon}.

Magnetic tunnel junctions (MTJs) with MgO barriers \cite{Yuasa04,Parkin04}
have great potential for applications in magnetic random access memory
elements and high-frequency generators \cite%
{Oh2009,Deac2008,Jung2010,Matsumoto2009}. Band structure calculations of
isomorphic Fe$|$MgO$|$Fe layered structures predicted a large drop in the
electric resistance when the relative magnetization direction of the two
ferromagnets switches from antiparallel to parallel \cite%
{Butler01,Mathon2001}. The subsequently observed large tunnel
magnetoresistance (TMR) \cite{Yuasa04,Parkin04} can be explained in terms of
the symmetry matching of only the majority-spin states in Fe with the $%
\triangle _{1}$-band of MgO, which is the by far least evanescent in the
gap. The tunneling ratio of the majority spin electrons is therefore
relatively high while minority spin states are efficiently filtered out by
the MgO\ barrier. However, a quantitative first-principles description of
transport in magnetic tunneling junctions is complicated by defects. The
chemical composition of the interface strongly affects the TMR \cite%
{DeTeresa99,miao2008,Mather2006} and various interfacial defects have been
identified to reduce the TMR \cite{Xu2006,Velev2007,ke10}. The I-V curves
alone cannot discriminate between the possible different origins that reduce
the TMR.

According to conventional wisdom shot noise in tunnel junctions is classical
\cite{Blanter}, in agreement with earlier experiments \cite%
{Guerrero2007,Sekiguchi}. Recent evidence that shot noise in MTJs is
suppressed in the parallel configuration came as a big surprise \cite%
{Arakawa}. In order to resolve this issue we present parameter-free
calculations of shot noise in magnetic tunnel junctions. We compute
sub-Poissonian shot noise for the parallel magnetic configuration and
explain the results in terms of highly-transmitting resonant tunneling
between states localized at the interfaces between the ferromagnet and
insulator. The agreement between first-principles theory and experiments
\cite{Arakawa} is quantitative when disorder is taken into account. These
results provide strong evidence of coherent transport and (additional) proof
for the very high quality of the MTJs used in that study.

According to the scattering theory of transport a two-terminal conductor
subjected to a sufficiently small bias voltage $V$ leads to a time-averaged
electric current
\begin{equation}
\langle I\rangle =\frac{e^{2}}{h}\left\vert eV\right\vert \mathrm{Tr}\left(
\mathbf{t}^{\dag }\mathbf{t}\right)
\end{equation}%
and shot noise%
\begin{equation}
S=\frac{2e^{2}}{h}\left\vert eV\right\vert \mathrm{Tr}\left( \mathbf{r}%
^{\dag }\mathbf{rt}^{\dag }\mathbf{t}\right)
\end{equation}%
where $\mathbf{t}$ and $\mathbf{r}$ are the matrices of the transmission and
reflection coefficients in the space of the transport channels of the leads
to the scattering region. These equations become more transparent by making
use of the distribution function $\rho \left( T\right) =\sum_{n}\delta
\left( T-T_{n}\right) $ of the eigenvalues $\left\{ T_{n}\right\} $ of the
transmission matrix $\mathbf{T}=\mathbf{t}^{\dag }\mathbf{t,}$ where $%
T_{n}\in \left[ 0,1\right] :$%
\begin{equation}
\langle I\rangle =\frac{e^{2}}{h}\left\vert eV\right\vert \int \rho (T)TdT
\label{I2}
\end{equation}%
\begin{equation}
S=\frac{2e^{2}}{h}\left\vert eV\right\vert \int \rho (T)T(1-T)dT=2eF\langle
I\rangle ,  \label{S2}
\end{equation}%
where $F\leq 1$ is the Fano factor. For a conventional tunnel junction
transmissions are small and $\rho \left( T\right) $ is substantial only for $%
T\ll 1.$ We then may disregard the $\symbol{126}T^{2}$ term in the integrand
of Eq. (\ref{S2}) and classical shot noise corresponding to $F\rightarrow 1$
is recovered. Clearly, a suppression of the shot noise that would correspond
to a Fano factor significantly smaller than unity requires that $\rho \left(
T\right) $ is significant at transmissions close to unity. Indeed, below we
find such highly transmitting states in MTJs with sufficiently thin barriers.

Previous theoretical treatments of the statistics of quantum transport have
been limited to simple models. While these can be sufficiently accurate for,
\textit{e.g.,} structures defined on a two-dimensional electron gas, the
details of the electronic structure are essential to understand (nearly)
ballistic MTJs \cite{Butler01,Mathon2001,Heiliger}. This Letter reports the
results of material-specific first principles calculations of the statistics
of transport in Fe$|$MgO$|$Fe magnetic tunnel junctions as a function of
magnetic configuration, voltage bias, and interface morphology and compare
theory with experiments by Arakawa \textit{et al}. \cite{Arakawa}, in
particular the suppression of the Fano factor for the parallel configuration.

\begin{figure}[h]
\includegraphics[width=8.5cm]{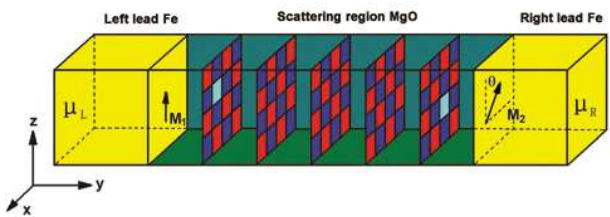}\newline
\caption{(Color online) Schematic diagram of a Fe$|$MgO$|$Fe(001) junction
containing 5 MgO monolayers. The magnetization \textbf{M}$_{1}$ of the left
lead coincides with the $z$-axis; the magnetization \textbf{M}$_{2}$ of the
right lead lies in the \textit{xz}-plane with angle $\protect\theta $. The
red and blue grids in the scattering region denote the O and Mg atoms,
respectively. We allow for interfacial disorder (oxygen vacancies, cyan
grid) at the Fe$|$MgO interfaces. The applied bias $eV=\protect\mu _{R}-%
\protect\mu _{L}$, where $\protect\mu _{R}$ and $\protect\mu _{L}$ are the
chemical potentials of the right and left leads, respectively.}
\label{scheme}
\end{figure}
We consider an MTJ consisting of a MgO barrier and two semi-infinite iron
leads as shown in Fig. \ref{scheme}. The electric current flows along the
(001) crystal growth direction. We incorporate the small lattice mismatch
between the leads and the barrier by a 3\% compression of the MgO lattice
constant. The self-consistent calculations are carried out with the
tight-binding linear muffin-tin orbital (TB-LMTO) \cite{Andersen86} surface
Green function method \cite{Turek97}. Disorder is treated using the layer
CPA (coherent potential approximation) \cite{Soven67}. The atomic sphere
(AS) potentials serve as input to the second step, in which the transmission
matrix is calculated using a TB-LMTO implementation \cite{Xia06}. Disorder
is modeled by large lateral supercells, distributing the self-consistently
calculated CPA-AS potentials randomly layer-for-layer in the appropriate
concentrations in as many configurations as necessary. Depending on the
defect concentration, most of our results are based on lateral supercells
containing 72 (two times $6\times 6$) or 128 (two times $8\times 8$) Fe
atoms per monolayer .

$\mathrm{TMR}\equiv R\left( AP\right) /R\left( P\right) -1$, where $R\left(
AP\right) \ \left[ R\left( P\right) \right] $ is the electric resistance for
the antiparallel (AP) [parallel (P)]\ configuration of the lead
magnetizations. For 5 MgO monolayers (L) at low bias $\mathrm{TMR}=3580\%$
for specular interfaces, which decreases drastically to 250\% when 5.56\%
oxygen vacancies (OV, the energetically most favorable defect) are
introduced at both interfaces. The TMR for the ideal junction is consistent
with published calculations \cite{Butler01}, while that for disordered
junctions is of the same order of magnitude as found in experiments \cite%
{Yuasa04}.

Based on the calculated scattering matrix, we compute the Fano factor for
various junction parameters. The results  are shown in Table \ref{layer_Fano}
together with the TMR.
\begin{table*}[tbph]
\caption{Barrier thickness dependence of the Fano factor in Fe$|n$MgO$|$Fe
MTJs for the parallel (P) and antiparallel (AP) configuration. The results
in square brackets are obtained for disordered junctions with 5.56\% oxygen
vacancies at the interfaces, where the error bar is given in round brackets.}
\label{layer_Fano}
\begin{center}
\begin{tabular*}{12.5cm}{@{\extracolsep{\fill}}ccccc}
\hline\hline
Fano factor & 3MgO & 4MgO & 5MgO & 7MgO \\ \hline
P & 0.64[0.65(2)] & 0.91[0.69(4)] & 0.97[0.87(4)] & 1.00[0.99(1)] \\
AP & 0.94[0.77(2)] & 1.00[0.94(1)] & 1.00[0.98(1)] & 1.00[0.99(1)] \\
TMR & 1320\%[165\%] & 2400\%[288\%] & 3580\%[250\%] & 5600\%[107\%] \\ \hline
\end{tabular*}%
\end{center}
\end{table*}
For thick barriers, the Fano factors are very close to unity, implying full
Poisson noise as expected. As the barrier gets thinner, the Fano factor of
the parallel configuration is increasingly suppressed. For a 5 MgO layers
junction with 5.56\% interfacial disorder, the Fano factor is $F_{P}=0.87(4)$
and $F_{AP}=0.98(1)$ for both configuration, close to the experimental
values $F_{P}=0.91(2)$ and $F_{AP}=0.98(1)$ for the same thickness \cite%
{Arakawa}. We can identify the majority ($\uparrow $) and minority ($%
\downarrow $) spin contributions to be $F_{P}^{\uparrow }=0.96$ and $%
F_{P}^{\downarrow }=0.72$, where $F_{P}^{\uparrow (\downarrow )}\equiv
S_{P}^{\uparrow (\downarrow )}/\left( 2e\langle I_{P}^{\uparrow (\downarrow
)}\rangle \right) $.

\begin{figure}[tbp]
\includegraphics[width=8.5cm]{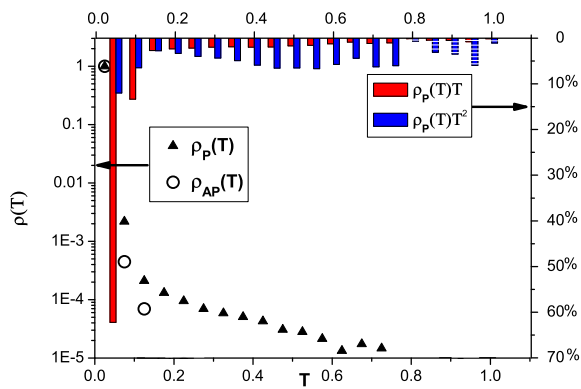}
\caption{Distribution of the transmission matrix eigenvalues $\protect\rho %
(T)$. The open circles are obtained for the AP configuration, for which
99.95\% of eigenvalues are $T_{n}<0.05$. Solid triangles represent results
for the P configuration. While most eigenvalues are still less than 0.05,
the few high values prove the presence of resonant tunneling states. The red
and blue histograms indicate the contribution of the integrands $T$ and $%
T^{2}$ to each eigenvalue interval (for P). Dashed bars indicated increased
statistical error due to the small number of $T_{n}$'s approaching unity.}
\label{PTN}
\end{figure}

In order to trace the origin of the shot noise suppression, we plot the
distribution functions of the transmission matrix eigenvalues $\rho
_{P/AP}(T)$ in Fig. \ref{PTN} involving $7\times 10^{6}$ eigenvalues over
the whole Brillouin zone. For P, we identify a few high values of $T_{n}$,
which according to Eqs. (\ref{I2},\ref{S2}) affect shot noise $S$ more
strongly than conductance $G$: 0.3\% of the eigenvalues are larger than 0.05
but contribute about 39\% to $G$ but 89\% to the integrand proportional to $%
T^{2}$ in Eq. (\ref{S2}), which suppresses $S$. The integrands proportional
to $T$ and $T^{2}$ are shown for each eigenvalue interval in the histograms
of Fig. \ref{PTN}. The dashed bars indicate a larger statistical error
caused by the small number of eigenvalues at high $T_{n}$. For AP, only very
few $T_{n}$ fall into the region between 0.05 and 0.1, the rest (99.95\%)
are all less than 0.05.

\begin{table*}[tbph]
\caption{OV concentration dependence of the Fano factor in 5MgO
MTJs for P and AP configurations. The impurity concentrations are
obtained by different
numbers of OVs at the interfaces per lateral unit cell. 2 OVs in the $%
6\times 6$ supercell correspond to 5.56\%, 3 or 4 OVs in an
$8\times 8$ supercell correspond to 4.69\% and 6.25\%
respectively. The conductances are given in units of
$10^{-5}e^{2}/h$ per Fe atomic interfacial area, where the
statistical error bar is given in round brackets. The conductances
in square brackets are obtained by the Green function method
\protect\cite{ke2008} as a check of the statistics of the
supercell method.}
\begin{tabular*}{17.0cm}{@{\extracolsep{\fill}}c|ccc|cc}
\hline\hline
Concentration &  & P &  & AP &  \\ \hline
& G(maj) & G(min) & Fano factor & G & Fano factor \\ \hline
0 & 68.00[68.50] & 3.47[3.51] & 0.97 & 1.95[1.95] & 1.00 \\
4.69\% & 89(3)[94.5] & 41(7)[33.6] & 0.85(3) & 35(3)[32.8] & 0.98(1) \\
5.56\% & 80(4)[91.0] & 44(10)[29.2] & 0.87(4) & 36(4)[37.5] & 0.98(1) \\
6.25\% & 79(3)[85.5] & 25(4)[26.9] & 0.95(2) & 38(3)[41.0] & 0.98(1) \\
\hline
\end{tabular*}%
\label{concentration_Fano}
\end{table*}
In Table \ref{layer_Fano} we can see that interfacial defects\ are necessary
to explain the observed shot noise suppression \cite{Arakawa}. The OV
concentration-dependent Fano factor for the 5MgO junction can be found in
Table \ref{concentration_Fano}. The statistics of our supercell calculation
is found to be good by comparison with a Green function formalism in which
impurity scattering is handled by the CPA \cite{ke2008}. Furthermore, even
though an OV concentration around 5\% suppresses the Fano factor, a further
increase leads to a dramatical recovery of the full shot noise. The full
shot noise observed in earlier experiments\cite{Guerrero2007,Sekiguchi} is
therefore consistent with higher disorder.
\begin{figure}[tbp]
\includegraphics[width=8.5cm]{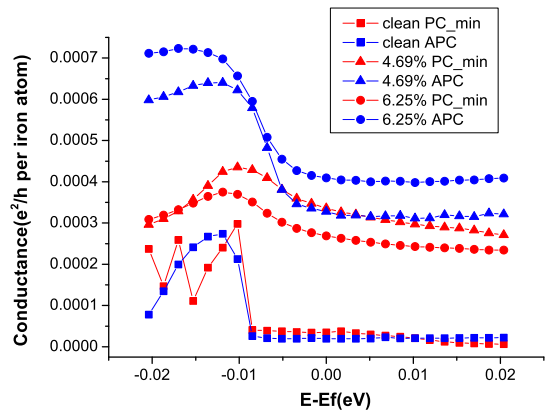}
\caption{Energy-dependent conductance for P (minority spin) and AP
configuration of a 5MgO MTJ for different impurity concentrations. Filled
squares represent the ballistic junction, triangles 4.69\% OV disorder, and
circles 6.25\% OV disorder. Red symbols stand for the minority spin
conductance in the P and blue ones for the conductance in the AP
configuration.}
\label{energy}
\end{figure}

In order to understand the sensitivity to the OVs, we plot the energy
dependence of the conductance of 5MgO MTJs with different OVs concentrations
in Fig. \ref{energy}. In ballistic junctions the minority spin conductance
for P and the AP conductance are strongly suppressed. Below the Fermi energy
high transmissions are observed, however. For P these are caused by the
overlap between interface states on both sides of the barrier. For AP, the
interface states exist only on one side of the barrier, but since their
symmetry is not orthogonal to the $\Delta _{1}$ states in the barrier and
the majority spin states on the another side, the conductance is still high.
Small amounts of OVs broaden and shift highly transmitting resonant channels
toward the Fermi energy, thereby suppressing $F_{P}$ (and the TMR). However,
a further increase of the disorder destroys the resonant channels thereby
recovering the full shot noise. The AP peak disappears and becomes a step
structure near the Fermi energy.

\begin{figure}[tbp]
\includegraphics[width=8.5cm]{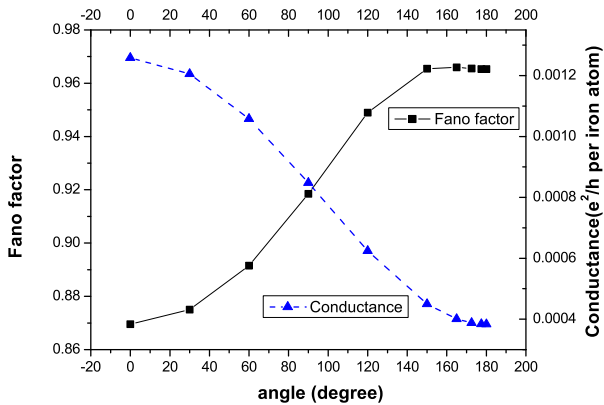}
\caption{Angular-dependent Fano factor (triangles) and conductance (squares,
per Fe atom at the interface) of 5MgO MTJs with 5.56\% OV disorder at the
interfaces. When $\protect\theta $ changes from P to AP, the conductance
decreases monotonously and Fano factor increases.}
\label{angle}
\end{figure}

A 5\% OV concentration appears to be close to the experiment \cite{Arakawa},
since it explains both Fano factors and the TMR. Further information may be
gained by the $\theta $-dependence of the Fano factor in Fig. \ref{angle}
for 5MgO junctions with 5.56\% OVs. $F\left( \theta \right) $ increases when
moving from P to AP which can be understood by the arguments above.

\begin{figure}[tbp]
\includegraphics[width=8.5cm]{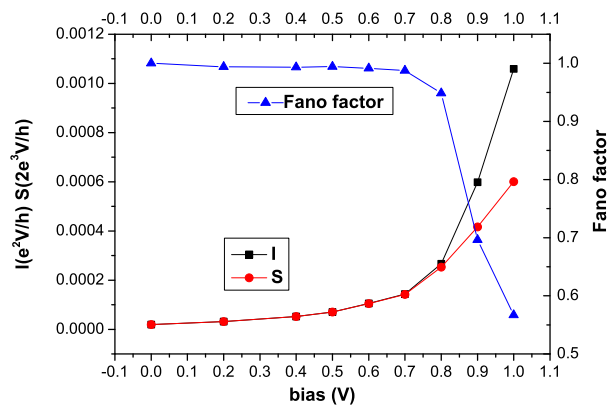}
\caption{Bias voltage-dependent Fano factor of an antiparallel $5$MgO MTJ. $I
$ and $S$ stand for current (per Fe atom at the interface) and shot noise
integrated over the bias window, respectively. }
\label{BIASF}
\end{figure}

An important issue of the current-induced spin transfer torque (STT) in MTJs
is its bias dependence. Recent experiments \cite{CWang2011} discovered a
nonlinear increase of the STT and current at an applied bias of 0.2$\,\unit{V%
}$. This value if far below the MgO band gap as calculated in the local
density approximation (LDA) \cite{Heiliger}. Since the LDA strongly
underestimates band gaps, the observed threshold must have a different
origin. A recent first principles analysis of the STT in Fe$|$MgO$|$Fe
junctions explained the threshold in terms of resonant transmission channels
in the AP configuration \cite{Jiabias}. This hypotheses can be tested by the
bias dependence of the shot noise. At finite bias $V$, the zero temperature
current and shot noise read \cite{Blanter}
\begin{eqnarray}
I\left( V\right)  &=&\frac{e^{2}}{h}\int^{eV}\left[ \int \rho (T,E)TdT\right]
dE \\
S\left( V\right)  &=&\frac{2e}{h}\int^{eV}\left[ \int \rho (T,E)T\left(
1-T\right) dT\right] dE
\end{eqnarray}%
Fig. \ref{BIASF} shows the integrated current and Fano factor as a function
of applied bias for an Fe$\left( \uparrow \right) |5$MgO$|$Fe$\left(
\downarrow \right) $ junction. The Fano factor is unity for low bias but
suddenly decreases with increasing bias at the threshold of the non-linear
current characteristic, which for a clean junction is at about 0.8$\,\unit{V}
$, consistent with the threshold bias in the STT. Small amounts of oxygen
vacancies in MgO can lower this threshold bias to become closer to the
experimental value \cite{Jiabias}. \textit{\ }

In conclusion, we compute sub-Poissonian shot noise for magnetic junctions
with thin MgO barriers from first principles in good agreement with
experiments. We interpret these results as strong evidence for resonant
tunneling states weakly broadened by disorder scattering. While it was known
that MgO based tunneling junctions can be grown with high crystalline
quality, we believe that the implied wave functions coherence over the
tunneling barrier is an important piece of information that lends
credibility to the prediction of large thermal spin transfer torques \cite%
{Jiathermaltorques}. This additional evidence of the superior electronic
properties of MgO junctions should have ramifications for the application of
this materials to other than ferromagnetic systems.

We gratefully acknowledge financial support from National Basic Research
Program of China under the grant No. 2011CB921803,2012CB921304, NSF-China,
the Dutch FOM Foundation, DFG Priority Program \textquotedblleft
SpinCat\textquotedblright , and EU-ICT-7 contract no. 257159 MACALO. We
thank the Authors of Ref.
\onlinecite{Arakawa}
for valuable discussions concerning their experimental results.

\end{document}